# Sensemaking of physics equations: Rationale, conceptualization, and core strategies

Julia Hofmann[1,*], Pascal Klein[1], Andreas Müller[2], Josefine Neuhaus[1,3]

[1] Faculty of Physics, Physics Education Research, University of Göttingen, Friedrich-Hund-Platz 1, 37077 Göttingen, Germany
[2] Faculty of Sciences, Department of Physics and Institute of Teacher Education, University of Geneva, Boulevard du Pont d'Arve 40, 1211 Genève, Switzerland
[3] Friedrich-Alexander-Universität Erlangen-Nürnberg, Freyeslebenstraße 1, 91058 Erlangen, Germany
[*] Corresponding author: julia.hofmann@uni-goettingen.de

**Abstract**
In physics instruction, equations are often treated primarily as tools for obtaining numerical results. Yet physicists also use them to describe dependencies, explain why those dependencies hold, predict what happens in new situations, test expressions against physical knowledge, and explore consequences that are not immediately apparent. Although such uses appear throughout the physics education literature, they have not been brought together in a systematic account, and the terminology used to describe them remains inconsistent. Drawing on research in mathematics education on symbol sense and the function concept, we distinguish six epistemic functions of physics equations: calculating, describing, explaining, predicting, evaluating, and exploring. We then characterize six strategies through which these functions can be realized: dimensional analysis, order of magnitude analysis, covariational analysis, special case analysis, limiting case analysis, and formal analogies, illustrating each with examples from introductory physics. To examine how these strategies are viewed within the discipline, we surveyed 31 physicists at one university faculty about five of the strategies. Respondents regarded the strategies as indispensable to physics, judged limiting case analysis to be more difficult for students than dimensional analysis, and reported that their own courses could devote more time to them. By distinguishing what physicists use equations for from the strategies by which they do so, the paper provides a vocabulary for sensemaking of physics equations and specifies what explicit instruction in these strategies involves.

# 1. Introduction

Physics condenses much of what it knows into equations. In a few symbols, an equation states how quantities depend on one another, embodies the principles from which that dependence follows, and reaches situations that have not yet been observed. Yet equations are frequently treated as mere tools for computing numerical values. In his inaugural lecture at King's College London in 1860, Maxwell set out a different aim for his teaching: "...we shall endeavour to put our calculations into such a form that every step may be capable of some physical interpretation, and thus we shall exercise powers far more useful than those of mere calculation – the application of principles and the interpretation of results" (Maxwell, 1990, p. 672). On this view, the physical sense of an equation does not lie on the surface of its symbols and has to be worked out.

Students, by contrast, often treat equations primarily as instruments for computation (Redish & Kuo, 2015; Uhden et al., 2012), substituting values without relating the result back to the situation, a pattern Tuminaro and Redish (2007) describe as plug-and-chug. Because physics equations are expressed in mathematical form, mathematics education research provides useful conceptual resources. In working on mathematical tasks, learners set aside what they know about the situation a task describes, a pattern Schoenfeld (1991) calls the *suspension of sensemaking*. In its best-known demonstration, primary school children were asked for the age of a captain whose boat carries 26 sheep and 10 goats. Most of them gave an answer by combining the two numbers, although the question could not be answered from them (Greer et al., 2002). The phenomenon has since been documented across age groups (Verschaffel et al., 2000). The concerns of physics and mathematics education research have long been intertwined. In an edited volume on informal reasoning and education, Schoenfeld (1991) subtitled his chapter "An informal attack on the unfortunate divorce of formal and informal mathematics". In the next chapter of the same volume, Clement (1991) examines how experts and science students reason with analogies and extreme cases, two of the strategies discussed later, and with physical intuition. Together, these strands of research suggest that making sense of equations requires more than formal manipulation: it involves bringing mathematical relations into contact with the situations, principles, and intuitions that give them physical sense.

Learners' reliance on computation has been linked to their view of what mathematics is for in physics: those who regard it solely as a computational tool tend to restrict their use of mathematics accordingly (de Ataíde & Greca, 2013). Moving beyond such a restricted use requires a precise account of what else an equation can be used for, and of how that is done. It requires, in other words, an account of what making sense of an equation involves. Sensemaking with equations has been conceptualized in physics education research (PER) and science education research (Gupta & Elby, 2011; Kuo et al., 2020; Odden & Russ, 2019; Zhao & Schuchardt, 2021), and strategies for making sense of equations have been examined empirically (Akinyemi et al., 2025; Warren, 2010; White et al., 2023). Several questions nonetheless remain open. Individual functions an equation can serve have been named (e.g., Airey et al., 2019; Palmgren & Rasa, 2024), but not systematically characterized. Terminology in this area is imprecise in two directions: the same term sometimes denotes different constructs, and the same construct appears under different names. Empirical work on the strategies has concentrated largely on evaluating whether an expression is plausible, whereas the same strategies can serve other purposes as well. That the strategies are characteristic of expert practice is commonly presupposed (Chasteen et al., 2012; White et al., 2023), but physicists themselves have rarely been asked directly (for an exception, see Lenz & Gire, 2016).

The present paper addresses these gaps. Its aim is to systematize the epistemic functions physics equations serve, to characterize the sensemaking strategies through which these functions are realized and to delimit them from neighboring constructs with recourse to mathematics education research (MER). Another aim is to examine how physicists involved in research and teaching judge the strategies' relevance and difficulty. These aims are pursued across the sections that follow: Section 2 takes up the concepts from mathematics education research that clarify how learners make sense of symbolic expressions, Sections 3 and 4 develop the epistemic functions and the strategies that realize them, and Section 5 reports an exploratory survey that complements this primarily conceptual contribution. The findings of the paper along with implications for teaching are discussed in Section 6.

Before proceeding, the following distinction is in order. In the literature relevant here, sense and meaning are often used interchangeably, almost as synonyms (*sensemaking*, *meaning-making*, *sensible*, *meaningful*). Yet meaning is also used in a different way, to denote the semantic or philosophical meaning of a term (Ernest, 1990; see Kilpatrick et al., 2005, for an in-depth discussion in mathematics education). Understood in this way, meaning is foundational to the sense at issue in sensemaking. The two terms should nonetheless be kept apart, and we use sense throughout this paper.

# 2. Foundations from mathematics education research

Making sense of an equation is not a concern of physics education alone. MER has long examined how learners make sense of symbolic expressions and of the relations they encode. This section is selective and takes up only those concepts on which Sections 3 and 4 draw.

## 2.1 Sensemaking and symbol sense

In MER, sensemaking is treated as one component of mathematical thinking alongside problem solving and metacognition (Schoenfeld, 1992). It rests on the conviction that mathematics is learned as part of a sensemaking activity rather than through the mastery of isolated pieces of knowledge, a conviction formulated against a teaching practice centered on routines and on the passive acquisition of content (Hiebert et al., 1997; Schoenfeld, 1992).

For symbolic expressions such as equations, a closely related construct is *symbol sense*, initially proposed by Fey (1990) and further developed by Arcavi (1994, 2005). Three of the abilities Fey lists recur in physics-specific form in Section 4. These are the abilities to compare orders of magnitude for functions of the form $n$, $n^2$, $n^3$, to inspect the result of an algebraic operation and judge whether it is likely to be correct, and to determine which of several equivalent forms is most appropriate for a given question. Sherin (2001) describes the counterpart for physics equations as a feel for expressions that guides what is done with them. One distinction within symbol sense is particularly important for the account developed below: manipulating a symbolic expression and reading through it to what it expresses are distinct, complementary activities of working with symbols (Arcavi, 1994, 2005), and competence in the former does not entail the latter.

## 2.2 *Grundvorstellungen* and functional reasoning

MER has also developed a framework for how such sense is grounded in the conceptual structure of a mathematical domain. *Grundvorstellungen*[1] are the fundamental ideas that give a mathematical concept its sense and guide its adequate use. They provide a way of relating the mathematical structure of a concept to learners' mental representations and to situations in which the concept is applied (vom Hofe, 1995; vom Hofe & Blum, 2016). The approach has been applied to quantity (vom Hofe & Blum, 2016), to equations and variables (Weigand et al., 2022), to differentiation (Greefrath et al., 2023), and to integration (Greefrath et al., 2021).

For functions, three *aspects* are distinguished (Doorman et al., 2012; Vollrath, 1989). On the *assignment aspect*, a function relates two variables and assigns a value to each argument. On the *covariation aspect*, it describes how changes in the argument and in the value relate to one another. On the *object aspect*, a function is regarded as a mathematical entity that can itself be characterized, transformed, and related to other functions. The assignment aspect foregrounds individual input–output relations, a *static* view, whereas the covariation aspect foregrounds coordinated change, a *dynamic* one. The three aspects bear on different parts of the sensemaking of physics equations treated in the following sections. Assignment underlies substituting a value into an expression, and thus calculation. Covariation underlies the examination of a dependency and, by extension, the approach to a limit. The object aspect underlies the recognition of an expression as an instance of a function type familiar from elsewhere.

[1] German term, literally "basic ideas", commonly rendered as "basic mental models" (Greefrath et al., 2023).

The covariation aspect is the one most directly relevant here and has been elaborated in its own right. Saldanha and Thompson (1998) describe covariational reasoning in terms of mentally coordinating two quantities as their values change in relation to one another. Carlson et al. (2002) further specify this reasoning as a hierarchy of mental actions, ranging from coordinating the values and directions of change of two variables, through attending to amounts of change, to reasoning about rates of change (for a review see Thompson & Carlson, 2017). Two properties of the construct matter for what follows. The changes that are coordinated need not be physically realized but may be imagined (Weigand & Günster, 2022), which is what allows the reasoning to reach configurations a system cannot occupy. And the independent variable need not be time, so that the coordination applies to any pair of covarying quantities. A widely used task illustrates this: learners imagine a bottle being filled and sketch the height of the water as a function of the volume of water poured in (Carlson et al., 2002).

### 2.3 What a physics-specific account adds

Concepts of this kind describe how a relation between variables is grasped, and PER has taken them up, in part under labels of its own (e.g., Redish, 2022; Zimmerman et al., 2020). They do not suffice, however, where an expression is read as the representation of a physical situation. Once an expression is embedded in physics, two additional layers of meaning come into play. Symbols stand for physical quantities, and the meanings associated with those quantities shape how the expression is interpreted (Redish & Kuo, 2015). In addition, a relation can be traced back to an underlying principle or mechanism, which is what distinguishes merely stating a dependency from accounting for it. Section 3 accordingly turns to how physics equations are made sense of and to the purposes for which they are employed.

## 3. Beyond the mathematical expression: Sensemaking and epistemic functions of physics equations

### 3.1 Making physical sense of mathematical expressions

The relationship between mathematics and physics is historically grown, complex, and mutually constitutive (Galili, 2018). Yet physics does not simply apply mathematics as it is practiced in the discipline of mathematics itself. Once embedded in physics, mathematics is put to a characteristically different use. As Redish (2006, p. 1) puts it, while mathematics "may be the language of science, [...] math-in-physics is a distinct dialect of that language." Physics equations are accordingly not merely symbolic expressions describing physical systems. They are condensed representations of physical knowledge, and physical sense is in part constructed through the interplay of mathematical structure and physical knowledge (Palmgren & Rasa, 2024; Redish & Kuo, 2015).

Making sense of symbolic expressions is part of the everyday repertoire of expert physicists, who routinely look beyond an equation's symbolic surface and thereby bring a mathematical expression to life. Chasteen et al. (2012) characterize this dynamic, flexible coordination of conceptual, physical, and mathematical knowledge as “thinking like a physicist”. In contrast, novices often struggle to make sense of equations in this way (Redish & Kuo, 2015; Uhden et al., 2012). What separates the two is not only how well an equation is understood, but also what it is taken to be good for. An expression that represents a physical situation can be read for the relation it states, drawn on to account for why that relation holds, or set against an expectation formed independently of it.

Odden and Russ (2019, p. 6) characterize *sensemaking* in science education as a dynamic process in which an explanation is built or revised in order to “figure something out”: a gap or an inconsistency has been noticed in one’s own understanding, and the mechanism behind the phenomenon is worked out so as to resolve it. It sets in where something is puzzling, unexpected, or in conflict with what one otherwise holds, and it concludes where the explanation constructed closes that gap. The account is not tied to any particular representation: its prototypical cases involve no formalism at all, and mathematical formalism appears in it as one of several representations a reasoner may coordinate.

A line of research in PER takes up precisely this case, where the representation coordinated is the formalism itself. *Mathematical sensemaking* denotes the search for coherence between formal mathematics and physics concepts, and the flexible translation between mathematical expressions and physical relations (Gupta & Elby, 2011; Kuo et al., 2020). Two directions of coherence seeking are distinguished within it: coherence within the mathematical formalism, and coherence between the formalism and the physical system it represents. The former corresponds to part of what MER describes as symbol sense (Section 2.1), and our interest lies in the latter. This is not simply the account by Odden and Russ (2019) specified for equations. Coherence of this kind is also sought as a matter of routine, where symbolic structure is coordinated with a physical relation without any explanation being constructed. The two accounts coincide wherever an expression is set against a physical expectation and an explanation is built to reconcile them (see Section 3.3.5). For the remainder of the paper, we use *sensemaking* to denote *mathematical sensemaking*, restricted to coherence between the formalism and the physical system it represents, and refer explicitly to Odden and Russ where the construction of an explanation or the resolution of an inconsistency is specifically at issue.

A similar division organizes the framework Zhao and Schuchardt (2021) propose on the basis of a review of the literature on instruction and on students’ problem solving. They distinguish a mathematics sensemaking dimension from a science sensemaking dimension. The latter rests on Odden and Russ’s definition and runs from labeling the variables as physical quantities, through the measure an equation supplies and the structural pattern that relates these quantities, to the causal mechanism that accounts for why the pattern holds. Where both dimensions are drawn on at once, they speak of *blended* sensemaking. Theirs is a systematic account of what is made sense of when an equation is read scientifically. It is built to categorize the sensemaking opportunities instruction affords and the types of sensemaking learners engage in, not the purposes for which a mathematical expression is taken up in a given situation.

To describe translation processes between mathematics and physics at various levels, modeling cycles have been proposed in mathematics and science education. The modeling cycle by Blum and Leiß (2006) in MER conceptualizes the process of transforming a real-world situation into a mathematical model, processing it mathematically, and interpreting and validating the results obtained. The real world and the mathematical world are treated as distinct domains in this cycle. From a PER perspective, mathematics and physics are fundamentally intertwined, and several authors have accordingly proposed modeling cycles that take this interaction into account.[2]

[2] For example: Redish & Kuo (2015) and Uhden et al. (2012). For a comprehensive overview of different models and approaches see Pospiech (2019).

Uhden et al. (2012) criticize Blum and Leiß's modeling cycle for neither accounting for varying degrees of mathematization nor distinguishing between the technical and structural roles mathematics can play.[3] Instead of separating the two domains, they introduce a *physical-mathematical model* in which physical concepts and mathematical representations coexist at different levels of mathematization. Within it, mathematization and interpretation constitute complementary, inverse processes: mathematization assigns mathematical structure to physical concepts, whereas interpretation gives sense to mathematical structures. Technical mathematical work such as algebraic manipulation takes place solely within the domain of pure mathematics, and the physical-mathematical model is connected to the rest of the world through simplification and validation. Uhden et al. (2012, p. 497) explicitly state that "qualitative physics is not a distinct model in itself" and locate it at the first level of the physical-mathematical model, without any mathematization.

This class of modeling cycles has been criticized more fundamentally for separating physical from mathematical reasoning and for the unidirectional character of its transitions. Current accounts call instead for a constant two-way mapping between the two domains (Sands, 2021; Tytler et al., 2013) and for a blending rather than a separation of formal and conceptual reasoning (Kuo et al., 2020; Sirnoorkar et al., 2023). The revised modeling cycle by Uhden et al. (2012) meets much of this: the two domains are merged, mathematization and interpretation are reciprocal movements along its levels, and its transitions are structural rather than chronological. What it retains is a detached domain of pure mathematics. Within the physical-mathematical model, an equation is one possible product of mathematization, representing a physical system or situation. What the modeling cycle accounts for is how an expression acquires physical sense and how it is validated against the situation. What the model does not distinguish is the range of purposes for which an expression, once available, can be taken up.

Whereas modeling cycles operate at the level of the process, Sherin's (2001) *symbolic forms* operate at the level of the individual expression: they are the elementary units of meaning that a reader attaches to the structure of a symbolic expression. A symbolic form combines a *symbol template*, a schematic pattern of symbols, with a *conceptual schema*, the intuitive idea associated with that pattern. Recognizing a pattern and linking it with a schema constructs a connection between mathematical structure and physical relationship (Airey et al., 2019). For instance, the template $[\ldots/\ldots x \ldots]$ is associated with the conceptual schema in which one quantity is seen as inversely proportional to another, as in the rearranged form of Newton's second law, $a = F/m$, where the acceleration decreases as the mass in the denominator increases (Sherin, 1996). Building on the symbolic forms, Sherin distinguishes classes of *interpretive devices* that describe how an expression is read: narrative, static, and special case classes[4]. As Karam (2014) notes, symbolic forms are closely tied to the structural role of mathematics in physics, since they link templates with conceptual schemata and thus with meaning. Symbolic forms specify which intuitive idea a reader attaches to a given structure. They do not specify what the reader is doing with the equation in attaching it.

Taken together, the accounts reviewed in this section characterize how a mathematical expression comes to make physical sense: through the coordination of mathematical expression and physical system, through movement along levels of mathematization, and through the pairing of symbolic structure with intuitive schemata. What they leave open is what an equation is then used for. This matters because the purpose steers the reading: which features of the same expression become relevant depends on whether one asks which quantities are related, why they are related in this way, or whether the relation can be trusted at all. Individual purposes are named throughout the literature, but in passing and within constructs built to answer other questions. Section 3.2 follows them through those constructs and Section 3.3 develops a systematic characterization of them as the *epistemic functions* of physics equations.

[3] The technical and structural roles of mathematics in physics according to Pietrocola (2008) are explicitly discussed in the following section.

[4] The special case device class relates an equation's behavior to extreme or restricted parameter values.

## 3.2 Epistemic concepts in research on equations

The following overview draws both on research on how learners engage with mathematical expressions and on analytical accounts of the roles mathematics plays in physics.

The learner's stance toward mathematics was examined by de Ataíde and Greca (2013). The authors investigated students' *epistemic views* of the role mathematics plays in physics, drawing among others on Hammer's (1994) work on epistemological beliefs, and identified three categories: mathematics as a computational *tool*, serving merely to produce numerical results; as a *translator*, expressing physical phenomena or structures; and as a *structure*, organizing physics concepts. A learner's epistemic view can directly determine how equations are approached in practice: students who regard mathematics solely as a computational tool tend to restrict their use of mathematics accordingly (de Ataíde & Greca, 2013; Redish et al., 1998; Sherin, 2006).

Domert et al. (2007) approach the learner's stance through *epistemological mindsets* toward understanding physics equations. They identify six components of what understanding an equation involves: recognizing symbols, recognizing the underlying physics, recognizing the equation's structure, linking the equation to everyday life, knowing how to use it, and knowing when to use it. Students typically draw on only a small subset of these. Applying an equation for calculation and recognizing symbols occurred most frequently, while understanding of structure was comparatively rare. This is a pattern the authors read as consistent with the reduction of equations to computational tools. Taken together with de Ataíde and Greca (2013), this suggests that a computational view of mathematics is both cause and consequence of a narrow repertoire of engagement with equations.

Within this group of studies based on students' self-reports, Airey et al. (2019) come closest to naming purposes directly. Asking 356 students how they know when they understand a physics equation, they identified ten themes, among which calculation, explanation, description, and prediction are recognizable as purposes (a part of the characterization of epistemic functions of equations proposed in Section 3.3). What the study records, however, are associations with equations as such, elicited independently of any situation in which an equation is put to use. The themes therefore document which purposes students name, but neither what distinguishes one from another nor what each demands of the reasoner.

Whereas the preceding studies rest on learners' self-reported understanding, a second line of research approaches this multiplicity of purposes through problem-solving behavior. Bing and Redish (2008, 2009) start from the observation that mathematics does several distinct things in physics at once: it mirrors physical relations, supplies a basis for calculation, links ideas into a web, and packages rules and previous results. They term this the "epistemic complexity" of mathematics in physics (Bing & Redish, 2009, p. 2) and take it as the starting point for investigating how learners manage these functions and how they frame their use of mathematics while solving problems.

They capture this with the concept of *epistemological framing*: a learner's often tacit answer to two questions, what kind of activity is going on in a given situation and what kind of knowledge counts as relevant (Bing & Redish, 2009, p. 4). Framing determines which epistemic resources a learner activates and can narrow the range of approaches considered legitimate in a given situation. Bing and Redish distinguish four such framings. In *invoking authority*, using mathematics means citing an established law, equation, or prior result. In *calculation*, it means carrying out an algorithmic or computational procedure. In *physical mapping*, an expression is read as corresponding to, and evaluated against, the physical situation it represents. In *mathematical consistency*, an equation is validated through its structural similarity to other, already trusted mathematical relations. A related account specifically of a problem-solving activity is offered by Tuminaro and Redish (2007), whose *epistemic games* describe the coherent sequences of moves by which learners navigate an episode of problem solving.[5]
Framings of this kind are not fixed characteristics of an equation but choices that learners make and reconsider while solving a problem. Only one of the four constitutes sensemaking as used here: *physical mapping* most directly captures the act of making physical sense of an equation, whereas the remaining three rest on trust in a source, on a calculational procedure, or on an analogy to another mathematical structure rather than on making physical sense of it. Physical mapping thus marks the region in which the functions of equations at issue here are located.

A further group of accounts asks not what learners do but what mathematics accomplishes in physics as a discipline. The distinction between a *technical* and a *structural* role of mathematics goes back to Pietrocola (2008) and is taken up by Karam (2014). The *technical role* casts mathematics as an instrumental, procedural tool, and a learner who attends exclusively to this dimension creates an artificial division between the physical and the mathematical world (Burkholder et al., 2023; Karam, 2014; Redish & Kuo, 2015). The *structural role*, by contrast, casts mathematics as an instrument for reasoning about physical phenomena, and from it several purposes can be derived: explanation through the physical interpretation of mathematical expressions, and evaluation of a result's consistency through the consideration of special or limiting cases, often accompanied by explicit metacognitive reflection on the reasoning process itself.

Palmgren and Rasa (2024) broaden this into a characterization of four roles that go beyond modeling and problem solving. A *syntactic* role organizes physical knowledge and the logical foundations of physical theories. A *representational and semantic* role expresses physical systems and relations in structured forms that gain meaning through their connection to physical concepts. A role of *reasoning and modification* supports the analysis and refinement of models. A role of *generating new information and explanations*, finally, yields novel physical insight and predictions for scenarios not yet examined. Palmgren et al. (2025) pursue the topic at greater breadth in a systematic review, extracting six roles mathematics plays in physics and physics education: it supports learning and achievement; enables mathematical manipulations; guides reasoning; facilitates experiments and modeling; serves as a language; and provides a structural foundation for physics as a science.

Karam and Krey (2015) add a further dimension, the *epistemological status* an equation holds. It is determined by asking where an equation comes from, what it represents, and why it is expressed the way it is, that is, whether one takes it to be a principle, a definition, an empirical regularity, or a derivation. The categories are not mutually exclusive, and the status can shift over the history of a field. Kepler's laws were first identified as empirical regularities and only later derived formally from Newton's law of gravitation. What an equation is used for may thus depend on what the reasoner takes its origin to be.

Table 1 summarizes the epistemic concepts invoked in this section, in the order in which they were introduced. To keep the overview readable, each entry cites one key reference. Functions surface in every one of these accounts, but none of them discusses the functions explicitly. Each is built around something else: a belief about mathematics, a type of knowledge activated in a situation, the place of mathematics in the discipline, or a mode of justification. What is missing is an account of the purposes themselves: which of them there are, how they are distinguished from one another, and what each demands of the reasoner. Section 3.3 develops such a systematic account.

[5] The concept of epistemic games will be further discussed in Section 4.

Table 1: Epistemic concepts relevant to the sensemaking of physics equations, in the order in which they are introduced.

| **Term** | **Explanation/Specification** |
|---|---|
| Epistemic view (de Ataíde & Greca, 2013) | The learner's conception of what mathematics is for in physics; constrains how equations are approached |
| Epistemological mindset (Domert et al., 2007) | Perceived critical attributes of what it means to understand an equation |
| Epistemological framing (Bing & Redish, 2009) | Tacit answer to what kind of activity is going on in a given situation; determines which knowledge counts as relevant |
| Epistemic game (Tuminaro & Redish, 2007) | A coherent sequence of moves through which a learner navigates an episode of problem solving |
| Epistemic roles of mathematics in physics[a] (Palmgren et al., 2025) | How mathematics is deployed in the discipline of physics |
| Epistemological status (Karam & Krey, 2015) | How an equation is justified within a theoretical reconstruction/the objective basis and kind of knowledge underlying an equation |
| Epistemic functions (this paper, Section 3.3) | What an equation contributes epistemically, i.e. the purpose it serves within sensemaking |

[a]The term "epistemic role" is not used by Palmgren et al. explicitly, but all the roles they discuss are related to thinking, understanding, and learning and thus fit well under the umbrella term of epistemic concepts as discussed here.

## 3.3 A systematic account of the epistemic functions of equations

Asking what a representation contributes epistemically is not peculiar to PER: in the philosophy of science, the epistemic functions of scientific models have likewise been characterized in a variety of ways without being consolidated into a single account (Poznic, 2017). We therefore propose a synthesis that draws on the theoretical accounts and on the empirical observations reviewed above. We distinguish six epistemic functions and characterize each. The synthesis is a reasoned selection rather than an exhaustive account, intended to structure the discussion, and the functions are not entirely disjoint.

*Definition (epistemic function)*: An epistemic function denotes the purpose for which an equation is employed in a given physical context, such as *calculating*, *describing*, *explaining*, *predicting*, *evaluating*, or *exploring*. In this sense, an equation is related to the physical situation it represents, rather than manipulated without reference to that situation.

Accordingly, an epistemic function is not a property of the equation itself, but of its use in a specific situation, constrained by the question posed to the equation and by the prior knowledge available to the reasoner. A given equation can therefore serve different functions.

The most basic function is calculating. Determining the value of a quantity by substituting numbers into an equation generates knowledge about the system. What falls outside the notion of an epistemic function is purely syntactic manipulation, that is, transforming one expression into another without relating the result to the physical situation it represents (Bing & Redish, 2008; Karam & Krey, 2015; Tuminaro & Redish, 2007).

Each function is characterized below and illustrated by an example. Table 2 summarizes the synthesis along with the literature in which a function is explicitly or implicitly addressed. The six functions differ in the question they answer and in what that question is directed at. *Calculating* is directed at a quantity of interest and yields a determinate value. *Describing*, *explaining*, and *predicting* are directed at the physical situation itself and yield a relation, a reason, or a consequence. *Evaluating* is directed at an expression or model and yields a judgment of its adequacy. *Exploring*, in contrast, is not directed at a predetermined kind of result. The functions are related rather than mutually exclusive, and a single episode of reasoning may address several of them.

Table 2: Epistemic functions of physics equations.

| **Epistemic function** | **Guiding question** | **Directed at** | **Literature** |
|---|---|---|---|
| Calculate | What is the value of a quantity of interest? | A physical quantity of interest | Bing & Redish (2009) |
| Describe | Which relation holds among the quantities involved? | Relationship between physical quantities | Bing & Redish (2009); de Ataíde & Greca (2013); Redish (2021a); Wolf et al. (2015) |
| Explain | Why does a relation hold? | The relation as it holds in the physical situation, and the principle or mechanism it is traced back to | Karam & Krey (2015); Palmgren & Rasa (2024) |
| Predict | What follows for a scenario not yet examined? | A scenario within the model | Airey et al. (2019); Palmgren & Rasa (2024); Redish (2017); Wolf et al. (2015) |
| Evaluate | Can the expression or model be relied upon? | The expression or model itself | Akinyemi (2021); Redish (2021b); Wilcox et al. (2013) |
| Explore | What follows from the expression itself? | Not fixed in advance | Galili (2018); Palmgren & Rasa (2024); Uhden et al. (2012) |

### *3.3.1 Calculate*

A basic epistemic function of an equation is the determination of a numerical value. Calculating answers the question of what value a quantity takes in a concrete situation (Bing & Redish, 2009). It thereby generates knowledge about the system, even where the result is a single number.
Take the force acting on each of two cables from which a street light is suspended symmetrically between two walls, $|\vec{F}|= \frac{mg}{2\cos\alpha}$, with $\alpha$ measured from the vertical (Hofmann et al., 2026). For a street light with mass $m = 5\ \mathrm{kg}$ and an angle of $\alpha = 75°$ from the vertical, the expression yields a force of approximately $95\ \mathrm{N}$ per cable, that is, roughly twice the weight of the object itself. This value is not contained in the expression as such and is only accessible by calculating it for a given scenario.

The value obtained in this way is bound to the configuration for which it was calculated. It says nothing about why the dependence takes this form, how the system behaves in other regimes, or whether the expression can be relied upon in the first place. Where the result is not related back to the situation the expression represents, calculating alone is not enough: it collapses into the purely formal manipulation that Tuminaro and Redish (2007) characterize as plug-and-chug. Calculating is in this sense the most basic and the most restricted of the functions considered here.

It is nonetheless indispensable. The strategies discussed in the following section draw on it whenever a limit is taken, a special case is examined, or an order of magnitude is estimated.

### 3.3.2 Describe

The most immediate epistemic function of an equation lies in reading its symbols as physical quantities and its structure as a relation holding between them. *Describing* in this sense answers the question of which quantities are involved and how they relate to one another (Bing & Redish, 2009; de Ataíde & Greca, 2013).

Take again the force on a cable of a street light suspended symmetrically between two walls: $|\vec{F}| = \frac{mg}{2\cos\alpha}$ (Section 3.3.1). The equation captures which quantities are relevant here and that the force depends on the angle: the more horizontal the cables, the greater the force on each cable (Section 3.3.3). The same system can be described equivalently by $|\vec{F}| = \frac{mgl}{2\sqrt{l^2-(d/2)^2}}$, expressed in terms of cable length $l$ and wall separation $d$ and derived along a different path. Both expressions are physically correct, and both describe the system. Which relation is addressed and which quantities are foregrounded depend on the descriptive perspective adopted.

### 3.3.3 Explain

An equation connected to physical principles can serve to account for why a relation holds as it does (Karam & Krey, 2015; Redish, 2017). *Explaining* thus goes beyond stating a relation: it answers why the relation holds by tracing it back to a principle or mechanism drawn on in addition to the expression. An explanation depends on which principle or mechanism one invokes, and on what the situation makes available.

Take again the street light suspended symmetrically between two walls, $|\vec{F}| = \frac{mg}{2\cos\alpha}$, with $\alpha$ measured from the vertical (Sections 3.3.1 and 3.3.2). Why does each cable have to withstand more than half the weight, and why does this force grow as the cables approach the horizontal? The expression, read together with the equilibrium of forces at the suspension point, accounts for this. A cable can only pull along its own direction, so of the force it exerts only the vertical component holds the lamp up, and that component is given by the cosine of the angle to the vertical. Since the arrangement is symmetric, the two cables pull with equal force at equal angles, so their horizontal components cancel at the street light and contribute nothing to carrying the load. What the two vertical components have to balance is the full weight, which is what the factor $2\cos\alpha$ in the denominator states. As the cables approach the horizontal, the vertical share of each cable force shrinks and the force required grows without bound. The same decomposition shows what the anchors have to hold: at the wall the horizontal component is not canceled, so each anchor takes a horizontal pull in addition to half the weight.

### 3.3.4 Predict

An equation taken to hold for a physical system can be carried forward to a case that has not yet been examined, provided the model is considered accurate for it (Airey et al., 2019; Palmgren & Rasa, 2024; Redish, 2017). *Predicting* answers the question of what follows for such a scenario, one that is specified in advance: a value to be expected at a given instant, the course a process will take, or the outcome of an arrangement not yet realized.

Take the velocity of an object falling under linear drag, $v(t) = g\tau + (v_0 - g\tau)e^{-t/\tau}$, derived from the equation of motion $m\dot{v} = mg - kv$.[6] Here $\tau = m/k$ is the time constant of the fall. Given the mass of a specific object, the linear drag coefficient $k$, and the initial velocity, the equation yields the velocity to be expected at any instant and thus how it develops over the course of the fall in a case not previously examined. Both are concrete consequences, obtained within the boundaries of the underlying model and without addressing why the velocity depends on time in this way. Such a consequence is testable, provided the scenario can be realized experimentally. Set against a measurement from a falling-body experiment, it becomes the basis for evaluating the model.

[6] The model assumes a drag force proportional to the velocity, which holds at low Reynolds numbers (Re$\leq$ 1).

Applied to a fall that has already been observed, the same expression would describe rather than predict.

### 3.3.5 Evaluate

An equation can itself be the object of assessment: its plausibility or correctness can be judged and its limitations probed (Palmgren & Rasa, 2024). This involves examining whether an expression is consistent with physical intuition, prior knowledge, or empirical observation (Akinyemi et al., 2025). *Evaluating* answers the question of whether the expression or model under consideration can be relied upon, for example, in terms of its dimensions or its behavior in specific scenarios (Akinyemi, 2021; Wilcox et al., 2013).

Of the six functions, evaluating presupposes the most: it requires an expectation about the situation, and that expectation must be obtained independently of the expression under scrutiny. Evaluating thereby relates to the preceding functions: the expectation it tests against is itself formed by describing or explaining, for instance. Where expectation and expression diverge, the discrepancy-driven account of sensemaking applies (Odden & Russ, 2019): an inconsistency is noticed and an explanation is constructed to resolve it.

Take the period of a simple pendulum, $T = 2\pi\sqrt{l/g}$, for a point mass on a massless string swinging at small amplitude. Since all bodies fall with the same acceleration irrespective of their mass, the period of the pendulum is not expected to depend on the mass. Additionally, a longer pendulum traverses a longer arc for the same angular displacement while the driving tangential acceleration remains unchanged, so it should swing more slowly. The expression meets both expectations. The mass does not appear in it, and the period grows with the length of the string. This supports the equation's plausibility without showing it to be fully correct, since any expression of the form $a\sqrt{l/g}$ would pass the same check, whatever the value of the dimensionless factor $a$.

### 3.3.6 Explore

Working through an expression can serve to examine the situation it represents (Uhden et al., 2012). *Exploring* answers the question of what follows from the expression itself. It proceeds from the expression and examines its consequences for the physical context, other contexts, or the expression, without a predetermined question guiding it. In this way it is possible to gain information that was not accessible before, whether as a new insight for a learner or as a scientific discovery (Palmgren & Rasa, 2024).

Exploring occupies a special position among the six functions. The other five each aim at a result whose type is specified in advance (a value, a relation, a reason, a consequence, a judgment). Exploring, too, is directed, but its outcome is not specified in advance. The process is open rather than arbitrary. It presupposes no expectation about the outcome but enough physical knowledge to make sense of the equation and to recognize the significance of what emerges. Exploration is thus not random manipulation but is guided by a trained sense of which variations, limits, or rearrangements might prove informative. Exploring is, however, not reserved for experts or for a particular structural form.

Take the range of a projectile launched at an angle $\theta$ to the horizontal, $R = v_0^2 \sin(2\theta)/g$. Explored without a determinate question, the expression shows that the range is greatest at $\theta = 45°$ and that complementary angles yield the same range. Exploration also occurs at the scale of the discipline, where an expression can disclose physics not previously known. Maxwell derived from his field equations that electromagnetic fields propagate as waves, and found, to his own surprise, that their speed agreed with the measured speed of light. He thereby identified light as electromagnetic radiation (Longair, 2015). In both cases information is gained that was not accessible beforehand.
Exploring can lead into the other functions. What is discovered may then be described as a relation or traced back to a principle, so that exploring often stands at the beginning of a sequence. Together with its openness to any expression, this is also why the section that follows does not treat it separately.

# 4. Making sense of physics equations: Strategies linking mathematical structure and physical concepts

Section 3 characterized epistemic functions as the purposes for which an equation can be employed. Across physics practice and instruction, certain patterns of working with an equation recur: a dimensional balance is checked, a dependency is traced, a case is substituted, a limit is taken, and an analogous system is invoked. Such patterns are the concrete forms in which these functions are realized, and they are the subject of this section.

*Definition (sensemaking strategy)*: A recurrent pattern of (mental) actions that takes a given symbolic expression as its object, interrogates its mathematical structure, and relates the result back to the physical situation the expression represents, thereby realizing one or more epistemic functions.

The strategies differ primarily in what they interrogate: the dimensions of the quantities involved, the magnitudes those quantities take on, a dependency between them, a specific case, a limiting process, or a second, structurally similar expression. These features organize the sections that follow. The physical knowledge a strategy draws on varies likewise. Since it varies both in kind and in extent, it is treated comparatively in Section 4.6.3. The definition requires the result to be related back to the situation, which mirrors the delimitation drawn in Section 3.3: an expression manipulated without reference to the situation it represents realizes no epistemic function.

Strategies of this kind are established in the literature, though under heterogeneous labels such as *sensemaking strategies* (Hahn et al., 2017; Lenz, 2020) or *quantitative reasoning* (Loverude, 2015). Akinyemi (2021, p. 46) characterizes a subset of them as strategies that “involve investigating the ability of the expression to describe the physical world” and uses the specific term *evaluation strategies*. Since our interest extends beyond evaluation, we adopt the broader term *sensemaking strategies*.

These strategies can be situated within the broader concept of *epistemic games* (Collins & Ferguson, 1993; Tuminaro & Redish, 2007): coherent sequences of moves through which a learner navigates an episode of problem solving. Tuminaro and Redish (2007) introduce the *mapping mathematics to meaning* game[7], in which physical sense is made of a mathematical expression. This is the converse of the *mapping meaning to mathematics* game in which an expression is constructed from a conceptual starting point. The strategies discussed below can accordingly be understood as moves through which *mapping mathematics to meaning* is carried out. The converse direction lies outside the scope of this section: the strategies set out from an expression that is already available and do not account for how it was derived.

The next sections present six such strategies (Figure 1): dimensional analysis, order of magnitude analysis, covariational analysis, special case and limiting case analysis, and formal analogies. They were selected because they are prominent in the literature and each applies across a broad range of equations and contexts. Note that the set presented here is not exhaustive. Each strategy is introduced in its own terms, and we then ask which epistemic functions its application can realize, and under which conditions. A single strategy may realize several functions, and which one is realized depends on the question put to the expression and on the physical knowledge available to the reasoner. Exploring and calculating are not treated separately in what follows. As set out in Section 3.3.6, *exploring* is a way of using the strategies rather than a result one of them yields: most strategies realize it wherever an expression is interrogated without a determinate question guiding the inspection. *Calculating*, in turn, underlies the strategies to varying degrees, since most of them involve calculational steps (Section 3.3.1).

[7] Tuminaro and Redish (2007) discussed six epistemic games that capture students’ problem-solving approaches.

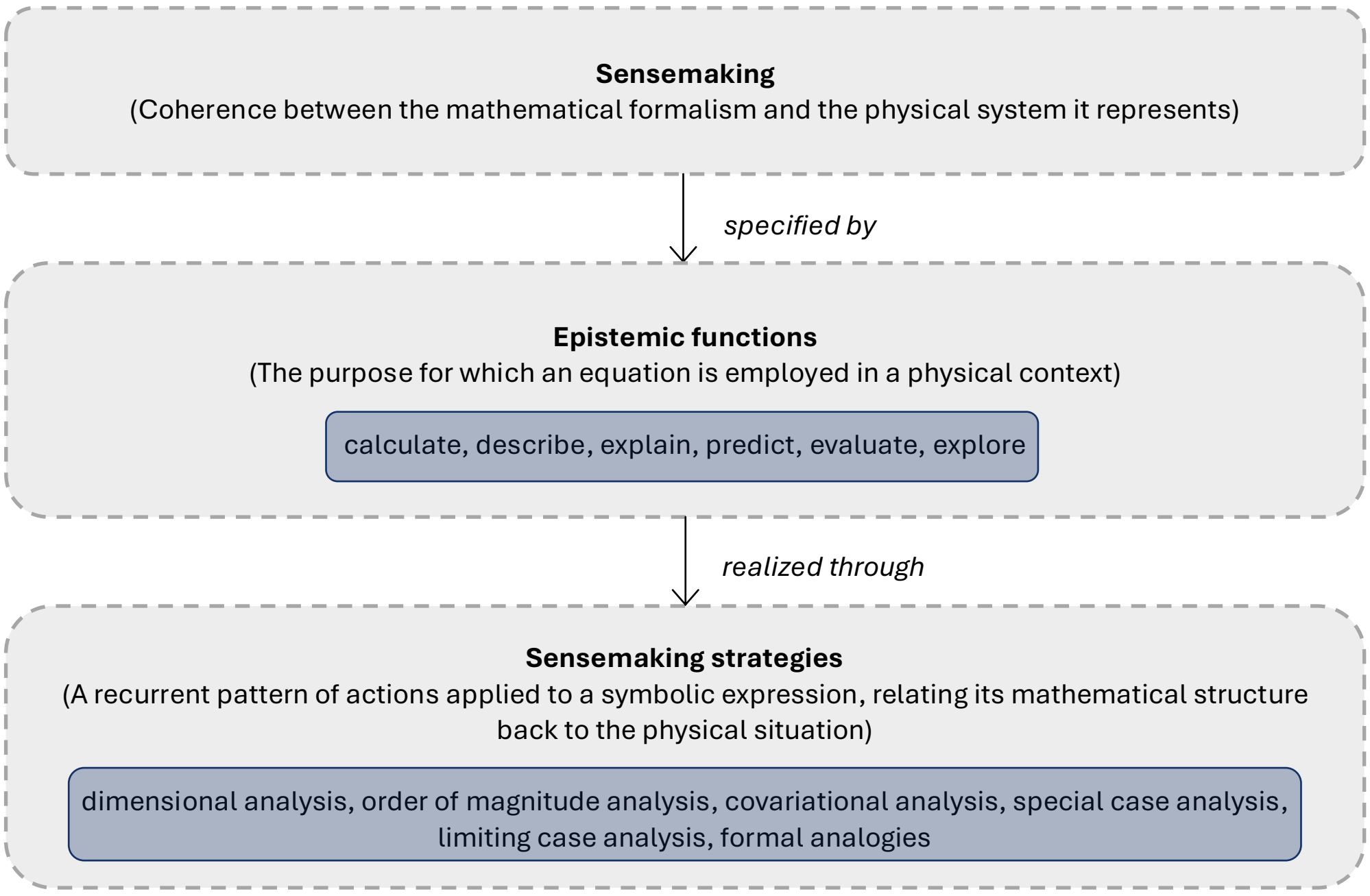

Figure 1: Sensemaking is specified by the epistemic function an equation serves in a given situation, and that function is realized through a sensemaking strategy. Sensemaking is used in the mathematical sense of Gupta and Elby (2011) and Kuo et al. (2020). The epistemic functions and sensemaking strategies shown are those characterized in Sections 3 and 4 and are not claimed to be exhaustive.

### 4.1 Dimensional analysis

Symbols in physics equations stand for physical quantities (Section 2.3), and a quantity is given not by a numerical value alone but by that value together with a unit (Bernstein & Friedman, 2009; Redish & Kuo, 2015). A statement such as $t = 30$ is physically empty unless a unit is given, which also specifies what kind of quantity $t$ denotes. Dimensional analysis (DiA) is the strategy that interrogates an expression at precisely this level.

A distinction is conventionally drawn between the *dimension* of a quantity (a fundamental property such as length L, time T, or mass M) and its *unit*, the concrete standard (e.g., meter m, second s, kilogram kg) by which that dimension is quantified within a given system of measurement, typically the *Système international d'unités*. In practice, however, the strategy is mostly carried out at the level of units, and it is presented that way in what follows.

Because quantities carry units, the *principle of dimensional homogeneity* holds: in a physically valid equation, the units on both sides must agree (Pescetti, 2008). Consider the buoyant force $F_B = \rho g V$ with displaced volume $V$ and density $\rho$ of the fluid. Substituting the units of the right-hand side yields

$$[\rho g V] = \frac{\mathrm{kg}}{\mathrm{m}^3} \cdot \frac{\mathrm{m}}{\mathrm{s}^2} \cdot \mathrm{m}^3 = \frac{\mathrm{kg\,m}}{\mathrm{s}^2},$$

which agrees with the unit of force on the left-hand side, $[\mathrm{F_B}] = \mathrm{N} = \mathrm{kg\,m}/s^2$. The equation is dimensionally homogeneous and, in this respect, physically plausible. This realizes the epistemic function *evaluate*. The expectation the equation is measured against comes from what the left-hand side denotes: $\mathrm{F_B}$ is a force, so the right-hand side must carry the unit of a force as well. What sets DiA apart from the other strategies is how little that expectation contains. Dimensional homogeneity is a necessary but not a sufficient condition for physical plausibility, and dimensionless factors such as $4\pi$ escape the check entirely.

DiA is not confined to checking an expression against itself. Where the units of a composite term are read as an indication of what kind of quantity that term represents, the strategy reaches beyond evaluation. Consider again the fall with linear drag (Section 3.3.4): $v(t) = g\tau + (v_0 - g\tau)\,e^{-t/\tau}$ with $\tau\ =\ m/k$. Since the drag coefficient carries the unit kg/s, the quotient $m/k$ carries the unit s and $\tau$ is therefore a time quantity. The same conclusion follows from the structure of the exponent: the argument of an exponential function must be dimensionless, so $t/\tau$ is only well formed if $\tau$ carries the dimension of time. Both routes identify $\tau$ as the time constant of the fall: mass and drag coefficient jointly set a time scale for the process, and they do so through their quotient. This *describes* a property of the situation.

## 4.2 Order of magnitude analysis

A physical quantity is given by a physical unit and a numerical value, and Section 4.1 was concerned with the first of these. The second carries information of its own about the situation. Whether a velocity is of the order of $10^{-2}\ \mathrm{m}/s$ or of $10^{8}\ \mathrm{m}/s$ decides whether a result is compatible with what is known about the situation. Order of magnitude analysis (OMA) interrogates an expression at this level and deliberately discards precision (in the values it uses and in the physics it carries) in order to determine the order of magnitude of a result.

In science and science education this way of thinking is established as *order-of-magnitude reasoning*: the use of approximate data and approximate calculations to arrive at estimates that are otherwise hard to come by, typically accurate to the nearest power of ten (Loretan et al., 2024). It appears under a range of further labels, among them *Fermi problems* (Edge & Dirks, 1983; Morrison, 1963), *back-of-the-envelope* estimation (Mahajan, 2018), and *guesstimation* (Weinstein & Adam, 2008). In their open form, Fermi problems require the relation to be constructed first, whereas what is treated here is the branch that takes a given expression as its object. Its standing in scientific practice is well documented, and Moore (1987, p. 5) points out that OMA is an “indicator, cultivator, and predictor of expertise”.

Consider a point source of radiant power $P$ radiating uniformly in all directions, and let $I$ denote the intensity of the radiation reaching a surface at distance $r$:

$$I = \frac{P}{4\pi r^2}\,.$$

Instantiated for the Sun, with a radiant power of order $P = 4 \cdot 10^{26}$ W distributed at the Earth’s distance of $r \approx 1.5 \cdot 10^{11}\ \mathrm{m} \approx 1\ \mathrm{AU}$[8], the expression yields an approximate value for the intensity of $1.4 \cdot 10^{3}\ \mathrm{W/m^2}$. Read as an order of magnitude, this value *describes* how the quantities relate to one another in this situation: a source of order $10^{26}$ W produces an intensity of order $10^{3}$ W/m$^2$ at a distance of order $10^{11}$ m. That magnitude can then be *evaluated* against an anchor of everyday experience: a square meter facing the midday Sun delivers roughly the power of a domestic toaster (1 kW). A rooftop photovoltaic module of about $2\ \mathrm{m}^2$, converting some 20 % of what it receives, accordingly delivers a few hundred watts. Commercial modules are rated around $400\ \mathrm{W}$, which matches the estimate.
Read as a physical quantity, the denominator is the area of the sphere of radius $r$ over which the entire power of the source is spread, undiminished. This *explains* the magnitude of the result: what separates $10^{26}\ \mathrm{W}$ from $10^{3}\ \mathrm{W/m^2}$ is not loss but dilution over the sphere's area of $3\cdot10^{23}\ \mathrm{m}^2$.
The same reading *predicts* the conditions at an orbit not yet examined. At the distance of Mars, about $2.3 \cdot 10^{11}\ \mathrm{m}$ (1.5 times the Earth's distance), the power of the source is spread over a sphere of about $7 \cdot 10^{23}\ \mathrm{m}^2$, and the intensity falls to about $600\ \mathrm{W/m^2}$. This is why solar-powered missions there carry correspondingly larger solar arrays.
The estimate rests on a model that sets parts of the physics aside. The expression gives the intensity above the atmosphere, whereas the kilowatt was read off at ground level, where absorption and scattering have removed roughly a quarter of it. The eccentricity of the Earth's orbit, which varies the distance over the year, was disregarded likewise. Which effects can be set aside without changing the order of magnitude has to be judged in advance, and that judgment is a central aspect of the strategy.

[8] The distance need not be looked up. Light takes about eight minutes to travel from the Sun to the Earth and moves at $3 \cdot 10^{8}\ \mathrm{m/s}$, so $s = vt \approx 3 \cdot 10^{8}\ \mathrm{m/s} \cdot 4.8 \cdot 10^{2}\ \mathrm{s} \approx 1.4 \cdot 10^{11}\ \mathrm{m}$, which suffices at this resolution. Approximate values an estimate requires can frequently be derived in this way from others more readily at hand.

### 4.3 Covariational analysis

An equation states how quantities depend on one another, in which direction and in what form. Covariational analysis (CoA) interrogates an expression at this level. Covariational reasoning, introduced in Section 2.2, is not confined to two quantities: S. R. Jones and Kuster (2021) situate it within the broader concept of *variational reasoning*, which also covers more complex forms of covariation such as *multivariation*, where more than two quantities vary in relation to one another. In what follows, covariational reasoning denotes the concept as characterized in Section 2.2 and CoA the strategy by which it is carried out on a given physical expression.

In physics, the dependency goes beyond a purely mathematical-functional one, since the quantities are bound to units and to a concrete physical situation. A common way of expressing it is scaling formulations of the *goes-like* type (Zimmerman et al., 2020). Neighboring concepts are addressed in the literature under labels such as *functional dependence* (Gifford et al., 2026; Hahn et al., 2017; Redish, 2022; White et al., 2023), *proportional reasoning* (Lenz, 2020; White Brahmia et al., 2021), *identifying dependencies or variables* (Burkholder et al., 2023; Loverude, 2015), and reasoning about the *sign* of a dependency (Sijmkens et al., 2026).

Consider again the intensity of a point source from Section 4.2: $I = \frac{P}{4\pi r^2}$.
The related quantities are $r$ and $I$, and the direction of the dependency can be read off first: as the distance grows, the intensity decreases. In this way the expression *describes* the situation. The exponent then tells how steeply it falls, which supports a *prediction* for a case not yet examined: moving a detector to twice the distance reduces the intensity it receives to a quarter, not to half as a linear reading would suggest. Expecting a reduction to half would be an instance of the *illusion of linearity* familiar from MER, which in physics is attenuated but persists as a default (De Bock et al., 2011).
Beyond that, one can *explain* why the dependency takes this shape. The radiation spreads uniformly, so at distance $r$ the entire radiant power passes through a spherical surface of area $4\pi r^2$. Because that area grows with the square of the radius, the power per unit area must fall as the inverse square of the radius. The relation is thereby traced back to a geometric argument.
This explanation, in turn, furnishes an expectation against which the expression can be checked, which realizes the epistemic function *evaluate*. The geometric argument yields an inverse-square dependence on distance, and the expression exhibits it. The geometric argument also settles the dimensionless factor $4\pi$, which a check on dimensions alone cannot reach.

## 4.4 Special case analysis and limiting case analysis

An equation represents a system in general, but it also applies to each specific scenario the system can be in. Some of those scenarios are particularly revealing, because in them the physics the equation encodes becomes easier to grasp. Special case analysis (SCA) and limiting case analysis (LCA) interrogate an expression by way of such scenarios.

Both strategies follow the same logic. A case is selected, and the expression is examined in that case, where it takes a form that can be related directly to the situation (Burkholder et al., 2020; Lenz et al., 2018; White et al., 2023). What matters is that the case be selected for what it can reveal. Substituting $\alpha = 27°$ into the expression for acceleration on an inclined plane discussed below yields a value. But nothing about the situation at $27°$ is known independently of the expression: whether the value is right could only be established by measurement. Applied in this way, the strategies collapse into *calculation*. Which cases are revealing must be judged from the situation the expression represents. We treat SCA and LCA as two distinct strategies, outlined below along with their formal difference.

### 4.4.1 Special case analysis

A special case is a value at which the structure of an expression simplifies, so that characteristic properties of the model and of the situation it represents become visible (Akinyemi et al., 2025; Lenz et al., 2018). Consider the acceleration of an object on an inclined plane with angle of inclination $\alpha$ relative to the horizontal and coefficient of friction $\mu$:

$$a = g(\sin\alpha - \mu\cos\alpha)\,.$$

Setting $\alpha = 90°$ describes the vertical configuration. Substituting gives $\sin\alpha = 1$ and $\cos\alpha = 0$, so the frictional term vanishes and the acceleration reduces to $a = g$. The case *describes* a relation the general expression does not display directly: for a vertical plane the acceleration is the gravitational acceleration alone, and the coefficient of friction no longer appears.

Why this holds can then be *explained*: the frictional force is proportional to the normal force, and on a vertical plane no component of the weight presses the object against the surface, so no normal force acts. This is an argument the expression carries in its structure, since the frictional term is proportional to $cos\ \alpha$ and thus to the normal force.

The explanation, in turn, furnishes an expectation against which the plausibility of the expression can be checked. An object not pressed against a surface is in free fall, so $a = g$ is what the model must yield here. The expression delivers exactly that, which realizes the epistemic function *evaluate*.

A different special case supports a *prediction*. Setting $\mu = 0$ removes the frictional term for every angle and leaves $a = g\sin\alpha$. For a plane on which friction is negligible, such as an air track, the acceleration at any inclination follows from this reduced expression, and it can be stated before the situation has been examined.

Special cases are not confined to numerical values. Setting one quantity equal to another specifies a case just as well. In the acceleration of an Atwood machine, $a = g(m_1 - m_2)/(m_1 + m_2)$, the case $m_1 = m_2$ yields $a = 0$, which agrees with the expectation that equal masses balance. What is substituted is then a relation between quantities rather than a value for one of them.

### 4.4.2 Limiting case analysis

In MER, the concept of a limit is understood to carry a static and a dynamic view, as the function concept does (Section 2.2; Fernández-Plaza & Simpson, 2016; Greefrath et al., 2016; Vollrath, 1989). The static view underlies the formal definition of a limit. The dynamic view captures the process of approach in which two changes must be tracked jointly. As a strategy, LCA takes a mathematical limit and examines how the expression behaves as a quantity approaches it (Sikorski et al., 2017). In the literature, this strategy is also referred to as *limiting behavior* (Chasteen et al., 2012; Wilcox et al., 2013), *extreme case reasoning* (Clement, 2009; Eichenlaub et al., 2016), or *evaluating limits* (Burkholder et al., 2020).
Consider again the fall with linear drag (Section 4.1):

$$v(t) = g\tau + (v_0 - g\tau)\, e^{-\frac{t}{\tau}},$$

and the behavior of the velocity for long fall times: $t \to \infty$. What happens in this case can be *explained*: the drag force grows with the velocity, so that it increasingly balances the weight, the acceleration tends to zero, and the velocity approaches a constant value. In the limit, the exponential term vanishes and $v$ approaches $g\tau = mg/k$. Expectation and expression agree, which realizes the epistemic function *evaluate*.
The limit also *describes* a property of the situation that the general expression does not display directly: the terminal velocity is set by mass, gravitational acceleration, and drag coefficient alone, while the initial velocity drops out entirely.
And it *predicts* the long-term outcome without the trajectory having to be followed. Whatever velocity the object starts with, it approaches $mg/k$.

#### *4.4.3 Relations between SCA, LCA, and CoA*

The three strategies SCA, LCA, and CoA are closely related, and the boundaries between them are drawn differently across the literature. The formal difference between SCA and LCA is visible in the same expression of the fall with linear drag. The special case $t = 0$ is carried out by substitution: $v(t = 0) = v_0$ returns the initial velocity. Whereas a special case is evaluated by substituting a specific value, a limiting case concerns a value that is approached but not attained within the model. Such limits frequently correspond to situations that are not physically realizable (Sokolowski, 2018). Nersessian (1993, p. 298) describes them in this sense as a manipulation of parameters within mental models, in which variables are varied "beyond what is physically possible". Much of the work on evaluation strategies uses special case analysis as the umbrella term, with limiting cases treated as one variety of it (Lenz et al., 2018; Warren, 2010), while work centered on limiting cases records special case among the labels used for the same strategy (White et al., 2023). We use them as two distinct strategies, as do other authors (Eichenlaub & Redish, 2019; Hahn et al., 2017).

Limiting case analysis also is closely related to covariational analysis. The dynamic view of the limit is structurally the same coordination of two changes that underlies covariation. Where the boundary between the two is drawn varies across the literature. Sokolowski (2021, p. 62) treats limiting case analysis as part of covariational reasoning, defining it as the "evaluation of the covariational expression by taking a limit". White et al. (2023) treat functional dependence as related to but distinct from limiting case analysis, Olsho et al. (2022) and Sijmkens et al. (2026) count limiting cases as one aspect of covariational reasoning, and Burkholder et al. (2020) list evaluating limits as a strategy in its own right. The proximity is reflected empirically: in student work on evaluating physics equations, the two strategies overlapped in both directions. Limiting cases were expressed qualitatively as dependencies between two quantities without a concrete extreme value being examined, and covariational reasoning was at times carried out in limit notation (Hofmann et al., 2026).

We nonetheless treat SCA, LCA, and CoA as three distinct strategies, because each of them interrogates something different. CoA examines a dependency across a range of values and asks how one quantity varies with another. SCA examines the expression at a single configuration and asks what it yields there. LCA examines the approach to a configuration and asks what the expression tends toward. The selection each requires differs accordingly: a dependency, a case, or a limit. In terms of the aspects of the function concept (Section 2.2), SCA draws on assignment, while CoA and LCA both draw on covariation, the latter in the dynamic sense of the limit.

## 4.5 Formal analogies

The same mathematical structure can recur across physically distinct domains. Where an expression shares its form with one that is already well understood, that understanding can be brought to bear on it.

Analogical reasoning denotes the process of identifying a situation structurally similar to a target situation and using the correspondence to transfer inferences from the better-understood case to the target (Clement, 2009; Podolefsky & Finkelstein, 2006). A formal analogy (FoA) establishes a connection between the structures of two equations that describe physically distinct phenomena (Gingras, 2015; Karam & Ricardo, 2012). As Gingras (2015, p. 530) puts it, two physically very different systems may nonetheless be "formally identical". FoA therefore always draws on a second symbolic expression, one that is structurally similar to the first. In terms of the aspects of the function concept (Section 2.2), formal analogies draw on the object aspect. The shared structure is recognized as a function type in its own right, understood independently of the quantities that instantiate it in either domain. This connects to the *symbolic forms* introduced in Section 3.1, where a recurring structural pattern (a symbol template) is paired with a conceptual schema (Sherin, 2001), though a formal analogy operates at the level of a whole expression rather than a single template.

Coulomb's law and Newton's law of gravitation are formally identical:

$$|\overrightarrow{F_C}| = \frac{|q_1 q_2|}{4\pi\varepsilon_0 r^2}, |\overrightarrow{F_G}| = \frac{G m_1 m_2}{r^2}.$$

Both describe a central force between two sources that scales with the product of the source quantities and falls off with the inverse square of their separation.

The analogy supports *describing* the electric interaction. Properties already established for gravity, such as its long range, carry over to the Coulomb force.

It also supports *explaining*. The inverse-square form of the gravitational law is accounted for by the spreading of the influence over a spherical surface[9] growing as $r^2$. What the analogy transfers is the explanatory pattern: because the two expressions share their form, the account given for one is available for the other.

The analogy further supports *predicting*. Consequences derived in one domain can be carried into the other before being established there. Newton's shell theorem states that a spherically symmetric mass distribution acts, seen from outside, like a point mass at its center. The formal analogy of the two laws implies its electrostatic counterpart: a spherically symmetric charge distribution acts, seen from outside, like a point charge at its center.

Conversely, the analogy furnishes an expectation and thereby supports *evaluating*. An expression for an electrostatic force falling off as $1/r^3$ would depart from the form its gravitational analogue takes, and that departure is a reason to examine the expression. What the analogy yields is an expectation, not a constraint: the two laws are logically independent, and the inverse-square form of the Coulomb force rests on its own empirical footing.

Analogies do not extend indefinitely, and where they end is itself informative. Charge, unlike mass, carries a sign, so the Coulomb force $\overrightarrow{F_C}$ may be repulsive as well as attractive, whereas gravity $\overrightarrow{F_G}$ is always attractive. Judging how far an analogy carries is part of the strategy rather than a limitation of it.

## 4.6 Similarities and differences between strategies

The six strategies were introduced separately, each in its own terms. Seen together, their application follows a common general structure. First, an object of examination is selected: a dependency, approximate values, a case, a limit, a set of dimensions, or a second expression. Second, the selected object is examined mathematically, for example by substituting or taking a limit. Third, the result is related back to the physical situation and read in its terms. The sequence is not confined to the mathematical or the physical level alone. Selection is guided by the structure of the expression as much as by what is known about the situation and thus draws on both the mathematical and the physical level. The second step is mostly technical and remains on the mathematical level. Relating back returns to the physical level, and the knowledge it draws on ranges from domain-specific physics to everyday experience and intuition. This sequence is not necessarily linear. For instance, where the result yields nothing informative, one selects a different case, another dependency, or an alternative strategy altogether. In expert practice, the mathematical operation and its physical reading are often not carried out one after the other but together, as a blended process (Kuo et al., 2013; Redish & Kuo, 2015).

[9] As explained in Section 4.3, using the example of radiation intensity at a given distance to illustrate CoA.

Within this common structure the strategies nonetheless differ, and Table 3 summarizes these differences. They differ in their procedure and in the object they interrogate, as the preceding sections have shown. Beyond this, they differ in how far the procedure is determined in advance, in how many opportunities for application an expression affords, and in the prior knowledge that application requires. Their range of realizable epistemic functions differs comparatively little: in the examples discussed, each strategy realized more than one function, dimensional analysis the fewest. The narrower range makes it no less important. It can be applied to any equation without further prerequisites, because dimensional homogeneity is a necessary condition that every physically valid expression must meet and checking it requires only the dimensions of the quantities and constants involved. The following paragraphs take up the openness of the procedure, the opportunities for application, and the knowledge each strategy draws on.

Table 3: Characteristics of the six sensemaking strategies: what each interrogates, by which operation, and how far its procedure is determined in advance.

| **Strategy** | **Object of interest** | **Operation** | **Openness** |
|---|---|---|---|
| DiA | Units of the quantities involved | Substituting dimensions or units | Mostly algorithmic |
| OMA | The orders of magnitude the quantities take on | Using approximate values; comparing scales | Open; the anchors and the admissible simplifications are selected |
| CoA | A dependency between two or more quantities | Varying quantities simultaneously | Open; the dependency is selected |
| SCA | A specific scenario | Substituting | Open; the case is selected |
| LCA | The behavior as a quantity approaches a limit | Taking a limit | Open; the limit is selected |
| FoA | The structure shared with a second expression from another domain | Matching structures | Open; the analogue is selected |

#### *4.6.1 Openness of the procedure*

The strategies differ in how far their application is determined in advance. Dimensional analysis is the least open strategy. For the homogeneity check, nothing needs to be selected beyond the relevant dimensions. This is what accounts for its narrower range of functions.

For the other strategies the procedure is less prescribed. What is examined must be selected: a dependency, a case, a limit among several possibilities, or a suitable second expression. Nersessian (1992), writing on mental models and thought experiments, emphasizes that the analysis of limiting cases follows no uniform algorithmic scheme, and White et al. (2023) extend this to special cases and to covariational reasoning applied to symbolic expressions.

The selection is consequential because not every admissible choice is informative. In the fall with linear drag (Section 3.3.4), the limit $g \to \infty$ is admissible and yields an infinite terminal velocity, a result that is plausible but discloses little about the model. Learners frequently choose cases of this kind, arriving at results that are correct without being revealing (Hahn et al., 2017). Which choice is informative cannot be read off the expression alone and must also be judged from the situation it represents, which is why the selection remains difficult and requires practice (White et al., 2023).

Order of magnitude analysis requires a selection of a different kind. What is chosen is not an object within the expression but the situation it is instantiated for, together with the level of approximation that situation admits: which values are close enough, and which effects may be set aside. The consequence of a poor choice differs accordingly. A poorly selected case or limit returns a result that is correct but uninformative, whereas an inadmissible simplification displaces the order of magnitude, and the estimate then fails by its own standard.

White et al. (2023) argue that this absence of algorithmic structure is not a deficiency of these strategies but the source of their value: it is what allows them to be applied flexibly and adaptively, in the way characteristic of expert practice.

### *4.6.2 Opportunities for application*

Beyond how far the procedure is fixed, the strategies differ in how many productive applications a given expression affords. The two are related but not the same: a procedure may be open while the expression offers few opportunities to apply it.

Here again, dimensional analysis offers the fewest opportunities. Within a given expression, the homogeneity check can be carried out only once. Order of magnitude analysis resembles dimensional analysis in being applicable to any expression, but for a different reason. Every expression can be instantiated with approximate values. What limits the strategy is not the expression but the availability of approximate values for the quantities it contains and of an anchor against which the result can be judged. Special case analysis and limiting case analysis fall in between. How many informative cases or limits an expression affords depends on how many quantities it contains and on how many of their values or limits correspond to revealing situations. The inclined plane offers $\alpha = 90°$ and $\mu = 0$, but not many more. Covariational analysis offers the widest range of possibilities. Every quantity in an expression depends in some way on the others, and each of these dependencies can be examined. The more quantities an equation relates, the more such applications it offers.

Formal analogies are not placed on this scale, because what limits them lies outside the expression. The question is not how many analogies a given equation affords but whether a structurally corresponding second equation exists at all, and whether the domain it comes from is understood well enough for anything to be transferred.

### *4.6.3 Prior knowledge*

Every application of a strategy draws on knowledge the reasoner brings to it. We use prior knowledge as an umbrella term for mathematical knowledge, physics content knowledge, conceptual understanding, physical intuition, and everyday experience (Eichenlaub et al., 2016; Eichenlaub & Redish, 2019; White et al., 2023). Which of these a strategy calls on is not fixed. It depends on the expression, on the situation it represents, and on what the person applying it already knows.

The strategies differ in the types of knowledge they draw on. Dimensional analysis draws mostly on knowledge of the dimensions of the quantities and constants involved. This is knowledge of facts, applied through a procedure that is largely automated (de Jong & Ferguson-Hessler, 1996). It calls for little knowledge of the situation the equation represents. The other strategies draw on knowledge of the situation itself: what happens in a selected case, how the system behaves at a limit, how one quantity responds when another changes, or how the system a second equation describes behaves. This is situational knowledge, and selecting what to examine additionally calls on strategic knowledge, which organizes how a problem is approached (de Jong & Ferguson-Hessler, 1996; Friege & Lind, 2006). Order of magnitude analysis draws on a further type of knowledge that the other strategies do not require: reference values for the quantities involved, held as anchor points that serve as conceptual benchmarks (M. G. Jones & Taylor, 2009). This knowledge often comes from everyday experience. We do not claim a complete assignment of knowledge types to strategies, because what an application calls for depends on the factors set out above.
This difference is reflected in what is known empirically (Hofmann et al., 2026; Sikorski et al., 2017; Warren, 2010; White et al., 2023). Dimensional analysis is acquired comparatively early and reliably, learners report high confidence in using it, it responds well to instruction, and difficulties arise less from its logic than from recalling dimensions. Covariational analysis, special case analysis, and limiting case analysis prove considerably more demanding and require sustained practice. Order of magnitude analysis is demanding for a different reason: its difficulties lie in the approximate values it must be supplied with, and in the technical steps it presupposes (Loretan et al., 2024).

#### *4.6.4 Multiple external representations*

Visualizations such as graphs and sketches likewise support sensemaking (Hahn et al., 2017; Tuminaro & Redish, 2007). They are treated here as a supporting technique rather than an autonomous strategy. Since an equation is itself an external representation, a symbolic one, visualizations add a further representation to the process, and the strategies above can be carried out with their support. This is consistent with accounts of learning with multiple external representations, according to which such representations complement one another, constrain interpretation, and support the construction of understanding (Ainsworth, 1999). A graph can, for instance, support explaining by making a covariational relationship between two quantities visible and thus easier to interpret (Sokolowski, 2021).

# 5. Perspectives from educational practice

## 5.1 Strategies in study and teaching practice

Sections 2, 3, and 4 proceeded from a theoretical standpoint. Whether the strategies are in fact present in study and teaching practice is a separate question, and the empirical picture is a mixed one. Students rarely apply them of their own accord, a pattern comparable to the suspension of sensemaking (Schoenfeld, 1991). Lenz et al. (2018) found special case analysis to be almost absent from students' reflection on their own solutions. Where students are explicitly asked to evaluate a result, they produce few strategies, and what they produce often remains computational or takes forms experts would not recognize as a strategy (Akinyemi et al., 2025). Where the strategies are taught explicitly, several studies report that students' use of them increases, though not to the same degree for each (Burkholder et al., 2020; Hofmann et al., 2026; Warren, 2010). Many authors conclude that the less algorithmic strategies should be integrated explicitly into instruction (Akinyemi et al., 2025; Burkholder et al., 2020; Lenz, 2020; Loverude, 2015; White et al., 2023).

As noted at the outset, the relevance of these strategies for expert practice is commonly presupposed rather than established. Where physicists have been asked, as in Lenz and Gire's (2016) interviews with faculty on how they use and teach dimensional analysis, the focus has been on a single strategy. The relevance attributed to the strategies as a group, for scientific work as much as for the curriculum, remains largely undocumented. The survey reported below is a first step toward closing this gap.

### 5.2 Expert perspectives on sensemaking strategies

To gain an initial insight into how physicists themselves regard such strategies, a survey was conducted at the Faculty of Physics at our university. Thirty-one faculty members, ranging from doctoral candidates to professors and all actively involved in teaching and research, participated in the online survey. Respondents rated five strategies (DiA, OMA, CoA, SCA, LCA) with respect to their significance for everyday scientific practice, their significance for physics education, the difficulty they present to students, and their integration into the respondents' own teaching. The questionnaire is provided in the Appendix. Ratings were given on a four-point scale and rescaled to the interval $[0,1]$, where 0 denotes no agreement and 1 full agreement. Order of magnitude analysis was part of the instrument, whereas formal analogies were not. The survey is exploratory rather than representative, since it draws on self-reports from a small sample at a single faculty.

The results present a consistent picture. Respondents affirm the relevance of the strategies for physics practice: they agree that the strategies play a role in their own scientific research ($M = 0.74, SD = 0.18$) and rate them even more strongly as an indispensable tool for physicists in general ($M = 0.86, SD = 0.14$). They attribute a comparable relevance to physics education ($M = 0.79, SD = 0.21$) and agree that the strategies should be introduced early in the degree program ($M = 0.83, SD = 0.20$). Across the five strategies, these ratings vary only slightly. All four statements were tested for differences between the strategies. Friedman tests indicate small but significant differences for three of them, namely the role in research, the standing as an indispensable tool, and the early introduction. The effect sizes are small throughout, and the Nemenyi post-hoc comparisons show no significant differences between individual strategies for any of the four statements. At the level of pairwise comparisons, none of the strategies is therefore set apart from the others.

The one aspect with a significant pairwise difference is the level of difficulty the strategies present to students. Limiting case analysis is judged comparatively demanding, dimensional analysis markedly less so (LCA: $M = 0.51, SD = 0.23$ vs. DiA: $M = 0.30, SD = 0.29$; Nemenyi post-hoc comparison between LCA and DiA: $p = 0.03$). This finding converges with what is known from empirical studies. It also runs in the direction the openness of the procedure would suggest: dimensional analysis, the strategy whose procedure is essentially fixed, is judged less demanding than limiting case analysis, one of the open ones.

When responding to questions concerning their own teaching, respondents see room for improvement in integrating all strategies into their courses ($M = 0.66, SD = 0.22$) while not regarding such integration as posing a fundamental difficulty ($M = 0.16, SD = 0.13$). The two items did not show significant differences regarding the individual strategies. Attributed relevance and reported implementation therefore diverge, and by the respondents' own assessment this is not due to any fundamental obstacle.

## 6. Discussion and conclusion

### 6.1 Contribution of the characterization

Making sense of physics equations builds on general mathematical competence, that is, on symbol sense and a grasp of the function concept, but it is not exhausted by that competence. Manipulating an expression and making sense of it are different achievements (Arcavi, 1994, 2005). The central contribution of this paper is a characterization of three constructs and of the relation between them, which has so far remained largely implicit (Figure 1). Sensemaking denotes the coordination of mathematical formalism with the physical situation an expression represents. The epistemic functions specify what an equation is used for within this process, whether to calculate, describe, explain, predict, evaluate, or explore. The sensemaking strategies are the recurrent patterns of action through which these functions are realized. The epistemic functions thereby add a dimension that existing accounts leave open. For instance, Zhao and Schuchardt (2021) categorize the types of sensemaking of equations in science along a science and a mathematics dimension, and symbolic forms specify which intuitive idea a reader attaches to a symbolic structure (Sherin, 2001). Neither specifies what an equation is used for in a given situation, and since the same symbolic form can serve different epistemic functions, the epistemic functions complement these accounts rather than replace them. Epistemological framing captures what kind of activity a learner takes to be going on in a problem-solving situation (Bing & Redish, 2009). Of the four framings, physical mapping is the one that constitutes sensemaking in the sense used here. It marks the region in which the epistemic functions are located, and the functions distinguish the uses an equation is put to within that region. The characterization is analytical rather than exhaustive. The functions are not entirely disjoint, and a single episode of reasoning may address several of them.

The characterization also responds to the terminological imprecision noted at the outset. The same term sometimes denotes different constructs. For example, in part of the literature, special case analysis serves as an umbrella term that includes limiting cases (Lenz et al., 2018; Warren, 2010), while elsewhere it is one of the labels for limiting case analysis itself (White et al., 2023). Conversely, the same construct appears under different names, as limiting case analysis does under limiting behavior, extreme case reasoning, and evaluating limits. A related problem concerns the boundary between constructs rather than terms. In PER, functional and covariational reasoning are often treated under a common set of labels without a sharp separation (Pospiech et al., 2019; White Brahmia et al., 2021; Zimmerman et al., 2020), whereas in MER covariation is one of three aspects of the function concept and thus narrower than functional reasoning as a whole (Section 2.2). Against this background, the paper defines epistemic function and sensemaking strategy explicitly and delimits the former from the neighboring epistemic concepts reviewed in Section 3.2. It further separates special case, limiting case, and covariational analysis, a boundary that is drawn differently across PER. This is not to prescribe a single usage. What matters is less that a particular term be adopted than that authors state which construct a term refers to and make the relevant distinctions and overlaps explicit. For the distinction of SCA, LCA, and CoA, MER supplies a conceptual basis, namely the aspects of the function concept set out in Section 2.2.

The relation between the two fields runs in both directions. MER contributes conceptual distinctions that PER has taken up only in part, among them the aspects of the function concept and the mental actions that structure covariational reasoning (Carlson et al., 2002; Olsho et al., 2022). PER, in turn, brings in what physics as a discipline supplies, namely the semantics of physical quantities: their dimensions, their plausible ranges of values, and criteria such as whether a result is physically admissible. It also grounds the strategies in expert practice, which the survey in Section 5 begins to document.

The examples used to illustrate the strategies in Section 4 show that they are not confined to evaluating a result, the use on which research has so far concentrated (Akinyemi et al., 2025; Sijmkens et al., 2026; Warren, 2010). Depending on the question put to an expression, the same strategy can also serve to describe, explain, predict or explore. The strategies discussed here were selected for their prominence in the literature, and the examples are taken largely from introductory physics. Neither the set of strategies nor the range of contexts is therefore claimed to be complete, and whether the strategies serve the same functions in more advanced domains remains to be examined.

The exploratory survey adds a first indication of how physicists involved in research and teaching judge the strategies. The respondents regarded the strategies as relevant to their own research and as an indispensable tool for physicists in general. Their relevance ratings did not set any one strategy apart from the others, which is consistent with treating them as a repertoire rather than as isolated techniques. Where the difficulty ratings differed, they did so in line with the openness of the procedure: limiting case analysis was judged more demanding than dimensional analysis, the least open of the strategies. This matches the empirical picture, in which dimensional analysis is acquired comparatively early and reliably while the more open strategies require sustained practice.

## 6.2 Implications for instruction

For instruction, the recommendation to make the strategies explicit can be specified in two respects. First, instruction should take into account how open a strategy's procedure is. Dimensional analysis can be introduced early and practiced as a largely fixed procedure, whereas the more open strategies, such as special case, limiting case, and covariational analysis, require practice in selecting what to examine and in judging which choices are revealing, since learners frequently choose cases that are admissible but uninformative (Hahn et al., 2017; White et al., 2023).

Second, the strategies need not be taught only as checks at the end of a solution. That use is valuable in its own right, since judging whether an expression can be relied upon connects the strategies to critical thinking. In science education, critical thinking has been conceived not as a generic skill but as thinking that meets the standards of the discipline in question (Bailin, 2002). This connection is likely to gain weight as artificial intelligence systems produce derivations and expressions that the user has not worked out and must nonetheless judge (Kortemeyer, 2023). Beyond that use, the guiding questions in Table 2 can be integrated into teaching and asked of a given expression: which relation it states, why that relation holds, what it implies for a scenario not yet examined, and what follows from the expression itself.

The design of tasks and the way they are used in teaching matter as well. MER does not trace the suspension of sensemaking to a lack of ability. Learners treat word problems as tasks in which the given numbers are to be operated on and the situation itself is not at issue, an expectation built up by the tasks they habitually meet (Greer et al., 2002; Verschaffel et al., 2000). In the terms of PER, what learners take the activity to be determines which knowledge they bring to bear (Bing & Redish, 2009). Tasks that can be completed by calculation alone are likely to reinforce a framing as calculation, whereas tasks that cannot be completed without relating an expression to the situation it represents call for sensemaking by design.

## 6.3 Conclusion

This paper set out to clarify what making sense of a physics equation involves. It distinguished three constructs: sensemaking as the coordination of mathematical formalism with the physical situation, the epistemic functions as what an equation is used for within this process, and the sensemaking strategies as the patterns of action through which these functions are realized. Concepts from MER grounded parts of this characterization. What the paper offers is an analytical framework rather than an exhaustive account, and with it a more precise vocabulary for describing, in research as well as in teaching, what learners are asked to do when they are expected to make sense of an equation. The physicists surveyed count the strategies among the indispensable tools of their discipline, yet they see room for integrating them more consistently into their teaching. Making the strategies explicit is therefore not an addition to physics instruction but a part of it.

# Appendix

Questionnaire of the exploratory online survey (Section 5.2). Faculty members of the Faculty of Physics ($N = 31$), rated five sensemaking strategies (DiA, OMA, CoA, SCA, LCA) on their significance for scientific practice and for physics education, their difficulty for students, and their integration into the respondents' own teaching.

| *Please indicate how much you agree with the statements relating to the different strategies on a four-point scale (1: "strongly disagree"; 4: "strongly agree").* *If you do not wish to make a statement, leave the corresponding cell blank. A brief explanation of the strategies can be found at the end of the questionnaire.* | **Dimensional Analysis** Example: Consideration of the units of both sides of a formula. Check that the units in the Bernoulli equation are correct and that the summands on the left-hand side of the equation actually correspond to pressures: $p + \frac{1}{2}\rho v^2 + \rho g h = \text{const. pressure}$. | | | | **Special Case Analysis** Example: Consideration of special or extreme values. Checking the trigonometric functions in the formula for acceleration on an inclined plane $a = g(\sin\varphi - \mu_k \cos\varphi)$ by considering particular angles, e.g.: $\varphi = 0°$ or $\varphi = 90°$. | | | | **Limiting Case Analysis** Example: Consideration of limiting cases. Checking the formula for the electric field strength of two point charges with a distance $d$ in $x$-direction: $\|\vec{E}\| = \frac{q}{4\pi\varepsilon_0}\left[\frac{1}{x^2} + \frac{x}{(x^2+d^2)^{3/2}}\right]$. Check whether $\|\vec{E}\|$ corresponds to a field strength of one point charge with a charge of $2q$ at infinite distance. | | | |
|---|---|---|---|---|---|---|---|---|---|---|---|---|
| | 1 | 2 | 3 | 4 | 1 | 2 | 3 | 4 | 1 | 2 | 3 | 4 |
| 0. I am familiar with this strategy. | ☐ | ☐ | ☐ | ☐ | ☐ | ☐ | ☐ | ☐ | ☐ | ☐ | ☐ | ☐ |
| 1. This strategy plays a role in my own scientific research. | ☐ | ☐ | ☐ | ☐ | ☐ | ☐ | ☐ | ☐ | ☐ | ☐ | ☐ | ☐ |
| 2. This strategy is an indispensable tool for physicists. | ☐ | ☐ | ☐ | ☐ | ☐ | ☐ | ☐ | ☐ | ☐ | ☐ | ☐ | ☐ |
| 3. This strategy should be introduced or taught early on in the study program. | ☐ | ☐ | ☐ | ☐ | ☐ | ☐ | ☐ | ☐ | ☐ | ☐ | ☐ | ☐ |
| 4. Mastering this strategy is necessary to follow advanced courses. | ☐ | ☐ | ☐ | ☐ | ☐ | ☐ | ☐ | ☐ | ☐ | ☐ | ☐ | ☐ |
| 5. This strategy leads to difficulties for students. | ☐ | ☐ | ☐ | ☐ | ☐ | ☐ | ☐ | ☐ | ☐ | ☐ | ☐ | ☐ |
| 6. This strategy enables students to gain a deeper understanding of physics concepts. | ☐ | ☐ | ☐ | ☐ | ☐ | ☐ | ☐ | ☐ | ☐ | ☐ | ☐ | ☐ |
| 7. This strategy should be practiced regularly by students. | ☐ | ☐ | ☐ | ☐ | ☐ | ☐ | ☐ | ☐ | ☐ | ☐ | ☐ | ☐ |
| 8. This strategy should be integrated into various physics courses. | ☐ | ☐ | ☐ | ☐ | ☐ | ☐ | ☐ | ☐ | ☐ | ☐ | ☐ | ☐ |
| 9. This strategy is an essential part of my own courses. | ☐ | ☐ | ☐ | ☐ | ☐ | ☐ | ☐ | ☐ | ☐ | ☐ | ☐ | ☐ |
| 10. Integrating this strategy into courses is difficult. | ☐ | ☐ | ☐ | ☐ | ☐ | ☐ | ☐ | ☐ | ☐ | ☐ | ☐ | ☐ |

| *Please indicate how much you agree with the statements relating to the different strategies on a four-point scale (1: "strongly disagree"; 4: "strongly agree").*<br><br>*If you do not wish to make a statement, leave the corresponding cell blank. A brief explanation of the strategies can be found at the end of the questionnaire.* | **Order of Magnitude Estiamtion**<br>Example: Checking the assertion that a melting of all glaciers will result in an expected sea level rise of around 80 m by estimating the order of magnitude of relevant parameters. Estimate the mean sea depth: 4000 m (minimum 1000 m; maximum 15000 m, geometric mean: 4000 m). Assume that about 1% of the earth's water volume is in the form of ice. An approximated rise of $4000\,\mathrm{m} \cdot 0,01 = 40\,\mathrm{m}$ corresponds to the order of magnitude of the assertion. | | | | **"Scales-like"-Consideration**<br>This strategy is generally understood to be the consideration of how two variables are related to each other. More precisely, it examines how a second variable changes when a first variable is changed. Example: Check the formula for the gravitational force: $F_G = G\frac{mM}{r^2}$. Consider whether the force decreases with increasing distance between the masses or whether $F_G$ "scales like" $1/r^2$. | | | |
|---|---|---|---|---|---|---|---|---|
| | 1 | 2 | 3 | 4 | 1 | 2 | 3 | 4 |
| 0. I am familiar with this strategy. | ☐ | ☐ | ☐ | ☐ | ☐ | ☐ | ☐ | ☐ |
| 1. This strategy plays a role in my own scientific research. | ☐ | ☐ | ☐ | ☐ | ☐ | ☐ | ☐ | ☐ |
| 2. This strategy is an indispensable tool for physicists. | ☐ | ☐ | ☐ | ☐ | ☐ | ☐ | ☐ | ☐ |
| 3. This strategy should be introduced or taught early on in the study program. | ☐ | ☐ | ☐ | ☐ | ☐ | ☐ | ☐ | ☐ |
| 4. Mastering this strategy is necessary to follow advanced courses. | ☐ | ☐ | ☐ | ☐ | ☐ | ☐ | ☐ | ☐ |
| 5. This strategy leads to difficulties for students. | ☐ | ☐ | ☐ | ☐ | ☐ | ☐ | ☐ | ☐ |
| 6. This strategy enables students to gain a deeper understanding of physics concepts. | ☐ | ☐ | ☐ | ☐ | ☐ | ☐ | ☐ | ☐ |
| 7. This strategy should be practiced regularly by students. | ☐ | ☐ | ☐ | ☐ | ☐ | ☐ | ☐ | ☐ |
| 8. This strategy should be integrated into various physics courses. | ☐ | ☐ | ☐ | ☐ | ☐ | ☐ | ☐ | ☐ |
| 9. This strategy is an essential part of my own courses. | ☐ | ☐ | ☐ | ☐ | ☐ | ☐ | ☐ | ☐ |
| 10. Integrating this strategy into courses is difficult. | ☐ | ☐ | ☐ | ☐ | ☐ | ☐ | ☐ | ☐ |